\documentclass[amssymb, preprintnumbers, showpacs, showkeys, aps, prb, superscriptaddress, twocolumn, longbibliography]{revtex4-2}

\usepackage{color} 
\usepackage{xcolor} 
\usepackage[colorlinks,bookmarks=false,citecolor=blue,linkcolor=blue,urlcolor=blue]{hyperref}
\usepackage{slashed}
\usepackage{graphicx}
\usepackage{amsmath}
\usepackage{latexsym}
\usepackage{epstopdf}
\usepackage{amsmath}
\usepackage{enumitem}
\usepackage{amssymb,amsmath}
\usepackage{multirow}
\usepackage{mathtools}
\usepackage{bbm}
\usepackage{bm}
\usepackage{amsfonts}
\usepackage{amssymb}
\usepackage{calligra}
\usepackage{calrsfs}
\usepackage{makecell}
\usepackage{comment}
\usepackage[normalem]{ulem}
\usepackage[USenglish,american]{babel}
\usepackage{subfigure}

\expandafter\ifx\csname package@font\endcsname\relax\else
 \expandafter\expandafter
 \expandafter\usepackage
 \expandafter\expandafter
 \expandafter{\csname package@font\endcsname}%
\fi
\begin{document}
\title{Effects of gravity on supersolid order in bubble-trapped bosons}
                
\author{Matteo Ciardi}
\email{matteo.ciardi@tuwien.ac.at}
\affiliation{Institute for Theoretical Physics, TU Wien, Wiedner Hauptstraße 8-10/136, 1040 Vienna, Austria}
\affiliation{INFN, Sezione di Firenze, I-50019, Sesto Fiorentino (FI), Italy}

\author{Fabio Cinti}
\email{fabio.cinti@unifi.it}
\affiliation{Dipartimento di Fisica e Astronomia, Universit\`a di Firenze, I-50019, Sesto Fiorentino (FI), Italy}
\affiliation{INFN, Sezione di Firenze, I-50019, Sesto Fiorentino (FI), Italy}
\affiliation{Department of Physics, University of Johannesburg, P.O. Box 524, Auckland Park 2006, South Africa}

\author{Giuseppe Pellicane}
\email{giuseppe.pellicane@unime.it}
\affiliation{Dipartimento di Scienze Biomediche, Odontoiatriche e delle Immagini Morfologiche e Funzionali, Universit\`a degli Studi di Messina, I-98125, Messina, Italy}
\affiliation{School of Chemistry and Physics, University of Kwazulu-Natal, 3209 Pietermaritzburg, South Africa}
\affiliation{National Institute of Theoretical and Computational Sciences (NIThECS), 3209 Pietermaritzburg, South Africa}

\author{Santi Prestipino}
\email{sprestipino@unime.it}
\affiliation{Dipartimento di Scienze Matematiche e Informatiche, Scienze Fisiche e Scienze della Terra, Universit\`a degli Studi di Messina, viale F. Stagno d'Alcontres 31, I-98166, Messina, Italy}

\begin{abstract}

Unveiling the principles behind self-organization in quantum systems is of paramount importance, both intrinsically and practically, in view of foreseeable technological applications.
Recently, increasing attention is being paid to atomic systems in curved geometries, which are a promising platform for the discovery of new emergent phenomena.
A notable example is that of a gas of ultracold atoms loaded into a thin spherical shell, according to a protocol introduced by Zobay and Garraway more than twenty years ago.
However, gravity prevents a dilute assembly of atoms from uniformly spreading throughout the shell, which explains why experiments on the condensation and superfluidity of bubble-trapped gases are usually conducted in space under microgravity conditions.
In this paper, we focus instead on strongly-interacting quantum particles in a bubble trap, choosing the cluster supersolid of soft-core bosons as testbed.
To study the impact of gravity on supersolid order, we consider a gedanken experiment in which the strength of gravity relative to the core repulsion is gradually enhanced.
Using path integral Monte Carlo simulations, we trace the parallel evolution of system structure and superfluidity at low temperature, finding that the latter is sizeable only when gravity is a small perturbation or, at the other extreme, so strong that particles are all gathered in one cluster at the bottom of the trap.
Finally, we assess the relevance of gravity for the equilibrium behavior of ultracold Rydberg-dressed atoms in a bubble trap, concluding that in some cases clues of the supersolid phase in the absence of gravity could be found even in a laboratory on Earth.

\end{abstract}

\date{\today}
\maketitle

\section{Introduction}
\label{introduction}

The problem of quantum particles in the presence of curvature has always attracted interest. Although basic three-dimensional shapes like spheres or cylinders are the backbone of common exercises in undergraduate quantum mechanics courses, they reduce to familiar solutions which hide the most subtle and general aspects of the problem. Beginning in the fifties, attempts were made to extend quantum mechanics to spaces and surfaces with intrinsic curvature, by modifying the Schrödinger equation to include a covariant derivative and an effective potential linked to curvature \cite{dewitt1957, jensen1971, dacosta1981, dacosta1982}. For a long time, these efforts remained theoretical, due to the lack of realistic experimental implementations. The last two decades have seen the advent of nanostructures, which can be modeled into various and increasingly exotic shapes, stimulating research into single-particle and transport properties of fermions in curved geometries \cite{neto2011, farajpour2018, dasilva2017}. Bosonic and many-body physics in curved geometries are also finally becoming experimentally accessible, thanks to developments in ultracold atoms.

The technical advances occurred during the '70s and '80s in the cooling and trapping of neutral atoms by light and static fields eventually culminated in 1995 with the first observation ever of Bose-Einstein condensation (BEC) in a quantum gas~\cite{anderson1995,davis1995}.
These important achievements, followed in subsequent years by constant refinements in the customization of atom-atom interactions (through, e.g., a clever use of magnetic Feshbach resonances), have significantly improved the experimental control over ultracold atoms, bringing it to a level of sophistication sufficient to engineer weakly- and strongly-interacting quantum systems of many sorts~\cite{Bloch2008,bloch2012, browaeys2020}, this way establishing atomic gases as a reliable platform for the development of quantum simulators~\cite{Feynman1982,Lloyd1996,Georgescu2014}.
In particular, by constraining atoms along one or two spatial directions~\cite{Petrov2004}, quantum many-body physics in low dimensions can be tested to an accuracy that was previously unattainable.

A major step forward in the manipulation of interactions was taken in 2001 when, by combining a quadrupolar magnetic field with a radio-frequency field, Zobay and Garraway demonstrated how to induce an adiabatic potential confining atoms in a thin-shell configuration~\cite{zobay2001}.
As long as gravity can be compensated or neglected, atoms will be uniformly distributed in the trap.
New impulse to study these shell-shaped (or ``bubble-trapped'') atomic gases came after the development, a few years ago, of microgravity facilities enabling, for the first time, the investigation of curved quantum systems not disturbed by gravity~\cite{lundblad2019,aveline2020,carollo2022}.
Curvature is a property of crucial interest to condensed matter physics, high-energy physics, cosmology, and even biology \cite{hagan2021,eckel2018,tononi2023,tononi2022,dubessy2024}, which is the reason why the possibility to introduce a curved constraint in ultracold atom setups has attracted so much attention.

From a theoretical point of view, there are at present open questions related to quantum many-body systems under curved confinement, which would deserve a more in-depth analysis.
For example, by a judicious choice of the confining surface, we could manage to guide the self-assembly of a many-body system and, in this way, engineer quantum gases which exhibit anomalous phase behaviors, like re-entrant melting, inverse melting, or liquid-liquid transitions, already demonstrated in classical fluids \cite{Tanaka2012,MendozaCoto2019,Neophytou2022}.
While the latter is an ambitious program, a more immediate task is to characterize superfluidity and supersolidity in a (finite) system of bosons confined in a closed surface.
In the paradigmatic case of a spherical surface, where crystalline order is frustrated, a supersolid phase is distinguished by a density field endowed with a non-trivial symmetry group (i.e., a finite point group), combined with a non-zero superfluid fraction. In less-symmetric cases, supersolids could simply be pictured as superfluids with a self-sustained density modulation~\cite{gross1957}.

The microgravity experiments conducted so far have been especially aimed at clarifying the features of BEC and superfluidity in {\em dilute} bubble-trapped gases, for which a number of theoretical studies already exist~\cite{tononi2019,bereta2019,diniz2020,rhyno2021,Tononi2024}, including dipolar condensates under weak gravity~\cite{Arazo_2021}. Conversely, significantly less is known about the phases of {\em strongly-interacting} atoms in a bubble trap, where, until recently, the only available predictions came from zero-temperature mean-field theory~\cite{prestipino2019a,PhysRevA.110.033322}.
That is why we have recently carried out extensive simulations of spinless bosons trapped in a spherical shell, using two different interactions, i.e., soft-core and dipolar~\cite{ciardi2024}.
Both potentials are known for providing stable supersolids in flat space at low temperature, where particles are grouped in clusters (or ``lumps'') of same size~\cite{macri2013,cinti2010,cinti2014a,tanzi2019,chomaz2019,bottcher2019,cinti2014b}.
For polarized dipolar bosons with realistic parameters, we have observed the formation of a supersolid ribbon of clusters wrapped around the sphere, perpendicularly to the direction of polarization.
In the case of penetrable bosons, we have identified a range of particle masses where the ground state is a cluster supersolid with polyhedral symmetry.
Clearly, the realization of supersolids on the sphere offers the chance to probe their physics in even more exotic conditions, e.g., in the presence of varying curvature or holes in the surface.

As anticipated, the distribution of particles in a bubble trap is perturbed by gravity, at least in the weak-interaction limit where particles are pushed to the bottom of the trap.
However, in a strongly-interacting system things may go different.
Indeed, the dipolar supersolid in Ref.~\cite{ciardi2024} is robust to Earth gravity, even though the clusters are now slightly shifted off the equator.
This unexpected outcome should be of inspiration to experimental physicists working with dipolar atoms, as it indicates that strong enough interactions can counterbalance the pull by gravity, allowing an easier investigation of the effects of curvature on quantum systems.

In this paper, we aim to assess the effect of gravity on the supersolid formed by soft-core bosons in a thin spherical shell.
To this purpose, we use path integral Monte Carlo simulations to investigate how the structure and superfluidity of the system evolve as the strength $g$ of gravity is gradually increased relative to the interparticle repulsion $\epsilon$.
As the original cluster arrangement is being undermined, an important issue will be the relationship between the residual clusterization and the degree of superfluidity, as measured by the superfluid fraction along the vertical direction and elucidated by the statistics of permutation cycles.
We shall see that clusters continue to exist when gravity becomes increasingly stronger, but their number is reduced progressively, until all particles condense in one superfluid cluster at the bottom of the trap.
Finally, by considering a realistic scenario in which the interaction between bosonic atoms arises from Rydberg dressing, we reach the conclusion that the supersolid phase of soft-core bosons can even be revealed in a laboratory on Earth.

The plan of the paper is as follows. After surveying in Section 2 the method employed in our study, we present and discuss our findings in Section 3. The final Section 4 is devoted to conclusions and outlook. Additional details on the more technical aspects of the simulation method are given in the appendices.

\section{Model and method}
\label{method}

Wishing to investigate the effect of gravity on an assembly of strongly-coupled quantum particles subject to a bubble-trap potential, our choice naturally goes to the supersolid phase of ultracold soft-core bosons, a system which we have recently characterized by quantum simulation~\cite{ciardi2024}.
Already within a mean-field framework ~\cite{prestipino2019a}, the stable high-density states of this system are clusters of particles forming a crystal-like structure, i.e., positioned at the vertices of a (semi)regular polyhedron.
The preference of soft-core bosons for clustered configurations is a purely classical phenomenon, promoted by the fat-Gaussian shape of the repulsive interparticle potential, as being illustrated, e.g., by the density-functional calculation presented in Ref.~\cite{prestipino2014}:
arranging the particles in equivalent clusters gives the system a distinct free-energy advantage at high pressure over diffuse partial overlap or the single-occupancy crystal of same structure.
Quantum indistinguishability adds a twist to the picture, since the emergence of clusters sufficiently close to each other brings about the possibility of a superfluid flow of particles, being more likely the smaller the particle mass.

In Ref.~\cite{ciardi2024}, we considered $N=120$ identical spinless bosons of mass $m$, pairwise interacting via the potential $\epsilon\theta(\sigma-r)$ (with $\epsilon,\sigma>0$), $r$ being the Euclidean distance.
The trap exerts an external potential on the particles, keeping them harmonically bound to a spherical surface of radius $R$.
We verified that, in a range of $R$ and $\lambda=\hbar^2/(2m)$ values, the thermodynamically stable phase is indeed supersolid at low enough temperature.

To simplify things further, we here assume an infinitely-sharp trapping potential, and thus consider soft-core bosons pinned to a spherical surface of radius $R$, now in the presence of gravity. In the following we call the gravity acceleration $a$, and $g = m a$. While this is in contrast to the common notation for the gravity acceleration, it improves the readability of our results. Then, gravity enters the calculation through an additional $g\sum_{i=1}^Nz_i$ term in the Hamiltonian, with $g>0$ and $-R<z_i<R$, $z_i$ being the $z$ coordinate of the $i$-th particle. 
Rather than assuming a specific $g$ value, we will consider a whole range of possibilities, so that we can obtain the full spectrum of responses of the Bose system to gravity.

For our numerical study we employ the path integral Monte Carlo (PIMC) method~\cite{ceperley1995}, which allows to compute the equilibrium properties of a bosonic system at finite temperature.
In a PIMC simulation, the generic configuration of the system is represented as a collection of closed ``polymers'' (world lines), each one consisting of the imaginary-time trajectories of a number of particles. To each quantum particle we associate a polymer of $M$ beads, corresponding to images of the particle on different slices of imaginary time. Polymers corresponding to separate particles can join together, forming a long polymer, which is identified by a``permutation cycle'', reflecting the order in which the component world lines are joined together.
The use of an ab initio Monte Carlo method, which exclusively relies on the form of the Hamiltonian and is exact up to numerical uncertainties, has strong theoretical foundations.
Therefore, our results are more reliable than those provided by simpler approaches, e.g., those based on the Gross-Pitaevskii equation, where the system is taken to be condensed from the very outset, or other theoretical studies relying on variational ansatzes.

For the simulation to run efficiently, it is essential to allow for a fast reshuffling of world lines.
To this purpose, we implement the worm algorithm~\cite{boninsegni2006b}, which efficiently samples bosonic permutations.
To deal with the spatial constraint, we resort to a modified version of the PIMC method which generates world-line configurations confined on the spherical surface~\cite{ciardi2024}, illustrated in full detail in Appendix~\ref{appa}.

To assess the superfluidity of the system, we measure its response to a slow axial rotation~\cite{leggett1970}.
In a supersolid, this occurs with a reduced moment of inertia relative to an ordinary solid of same mass.
The superfluid fraction $f_s$, namely the relative reduction in the moment of inertia, can be accurately estimated in a PIMC simulation by sampling the so-called “projected-area estimator”~\cite{sindzingre1989,ciardi2024}. Appendix~\ref{appb} describes the estimator for $f_s$ used in this paper.

For reasons of symmetry, we only inspect the superfluid fraction along three orthogonal axes.
For the isotropic soft-core fluid, the superfluid fraction along any of the principal directions is the same within errors.
Conversely, for $g>0$ the anisotropy of the interaction makes $f_s^{(z)}$ different from $f_s^{(x)}$ and $f_s^{(y)}$ over a wide range of parameters; therefore, in figures their values are reported separately. 

A powerful tool for analyzing the effects of quantum statistics in a PIMC simulation is the distribution of cycle lengths, reflecting the extent of quantum coherence in the system~\cite{feynman2010,jain2011a}.
Specifically, we consider the probability $P(L)$ to find permutation cycles made up of $L$ polymers, with $1\le L\le N$.
This probability is obtained by periodically updating the histogram of $L$.
In the superfluid regime we find permutation cycles encompassing any number of particles.
As the kinetic energy decreases, the system may enter a supersolid regime;
in this case, we still observe permutation cycles involving a large number of particles, though in lesser amount than in a superfluid, but still implying coherence among the clusters.
Finally, for smaller $\lambda$ values the clustered system becomes a normal (cluster) solid.
Here, the only permutation cycles occurring with significant frequency are those involving bosons within the same cluster, and quantum effects are thus limited to the single cluster.

\section{Results} \label{results}

\begin{figure*}[t!]
\begin{center}
\includegraphics[width=0.9\linewidth]{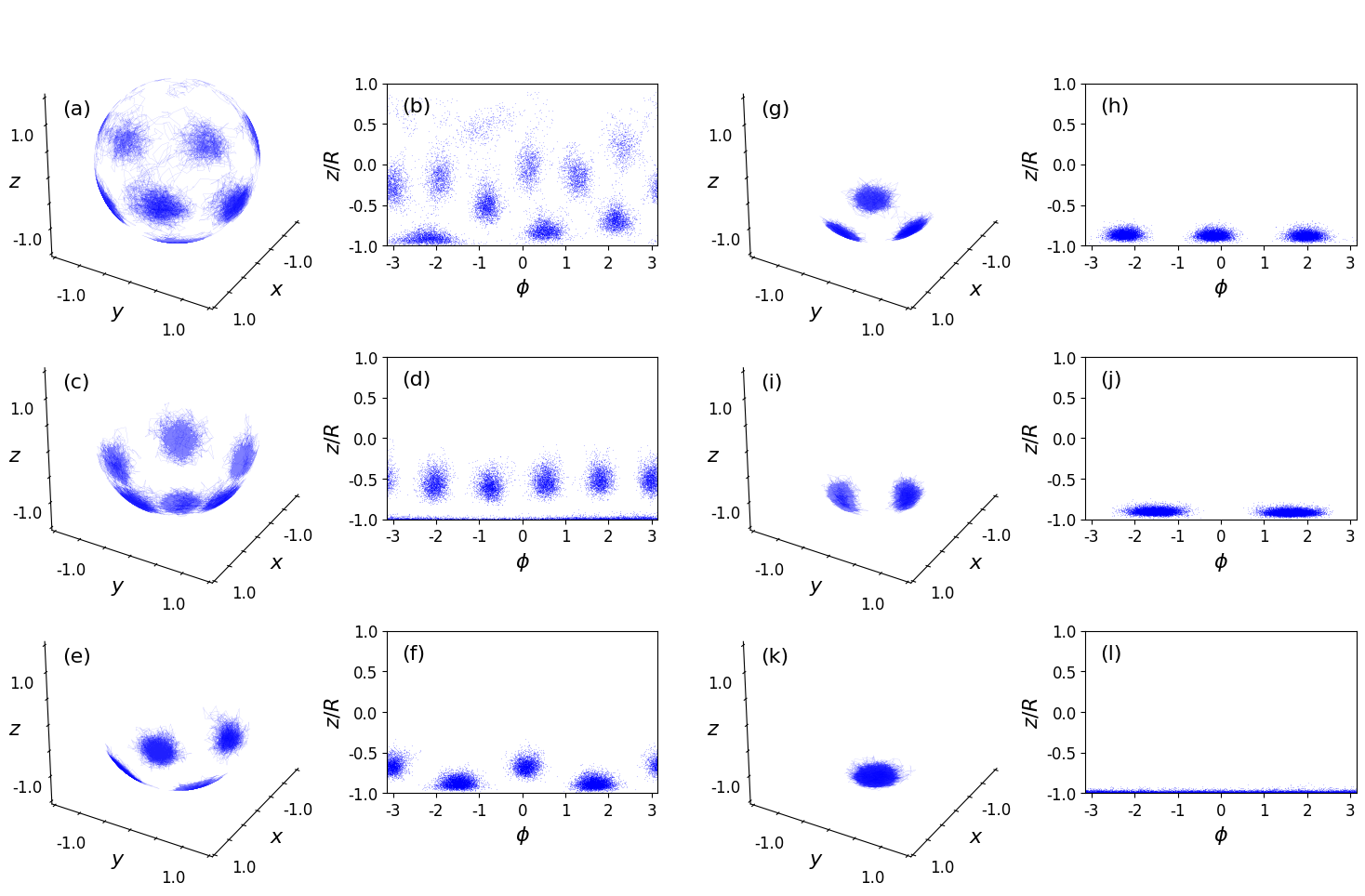}
\caption[]{Typical configurations of soft-core bosons on a sphere ($N=120,R=1.4,\lambda=0.16$, and $T=0.2$) for increasing gravity: (a---b) $g$=10, (c---d) $g$=20, (e---f) $g$=50, (g---h) $g$=150, (i---j) $g$=200, (k---l) $g$=220. Each blue dot represents a polymer bead (see text).}
\label{fig:conf_vs_gu}
\end{center}    
\end{figure*}

We hereafter present data for a system of $N=120$ soft-core bosons, i.e., sufficiently large to exhibit an emergent behavior.
In the following, all lengths are given in units of $\sigma$, and all energies are given in units of $\epsilon$. Temperatures are in units of $\epsilon / k_B$, where $k_B$ is the Boltzmann constant.

The scaling of $g$ is a delicate point, as we need to take into account the typical length scales and particle masses used in experiments. For example, we can consider a density of 4 $\mu$m$^{-2}$ on the spherical surface, and take the mass of a Rubidium atom $m \approx 87 \times 1.67 \times 10^{-27}$ kg. In this case, Earth gravity at 9.8 m/s$^2$ corresponds to $g \approx 8$ in simulation units, see more in Appendix~\ref{appc}.

Choosing $R=1.4,\lambda=0.16$, and $T=0.2$, in the absence of gravity ($g=0$) the system is supersolid:
the clusters lie at the vertices of a regular icosahedron and the superfluid fraction is $f_s\approx 0.38$.
Then, we increase $g$ gradually, while keeping the other parameters fixed, waiting at each step until the structure stabilizes before recording the values of $f_s$ and $P(L)$.

In Fig.~\ref{fig:conf_vs_gu} we show typical equilibrium configurations for a number of representative $g$ values.
In panel a ($g=10$), gravity is still weak enough that traces of the icosahedral structure are visible, although the bosons originally hosted in two clusters near the north pole are ``redistributed'' among the remaining clusters.
Looking at the spatial distribution of polymer beads in panel b, we see that particles are delocalized throughout the sphere, especially in the southern emisphere.
Understanding quantitatively whether the system is supersolid requires to look at the superfluid fraction.

When $g$ reaches 20 (panels c and d), the cluster structure has completely changed, now exhibiting a flower shape with a cluster at the bottom of the sphere surrounded by five ``petals''.
The space between the clusters is almost free of polymer beads, suggesting that bosons are localized.
It is important to stress that the ``structural transition'' occurring between $g=10$ and 20 is essentially classical, since the same transition is observed in a system of distinguishable quantum particles.
From this evidence, we argue that it is the structural transition that drives the ``quantum transition'' from particle delocalization to localization, and not vice versa.

As $g$ keeps on growing, the particles are pushed even further down on the surface and the number of clusters is progressively reduced (panels from e to j), until all bosons are confined in a single cluster centered at the south pole.
This latter case is illustrated in panels k and l, corresponding to $g=220$.
Since the temperature is very low, the coexistence of all particles in the same cluster suggests that the system is now superfluid.

We note that clusters containing a larger number of particles are not bigger in size.
Indeed, the cluster radius is essentially dictated by the range $\sigma$ of the repulsive barrier.
On the other hand, the more particles are hosted in a cluster, the more convenient is for those particles to stay in place, with the result that the cluster boundary becomes sharper.

More insight on the degree of quantum coherence in the system can be got from the probability $P(L)$ to find polymer chains (i.e., permutation cycles) involving $L$ particles.

\begin{figure}[t]
\centering
\includegraphics[width=1.\linewidth]{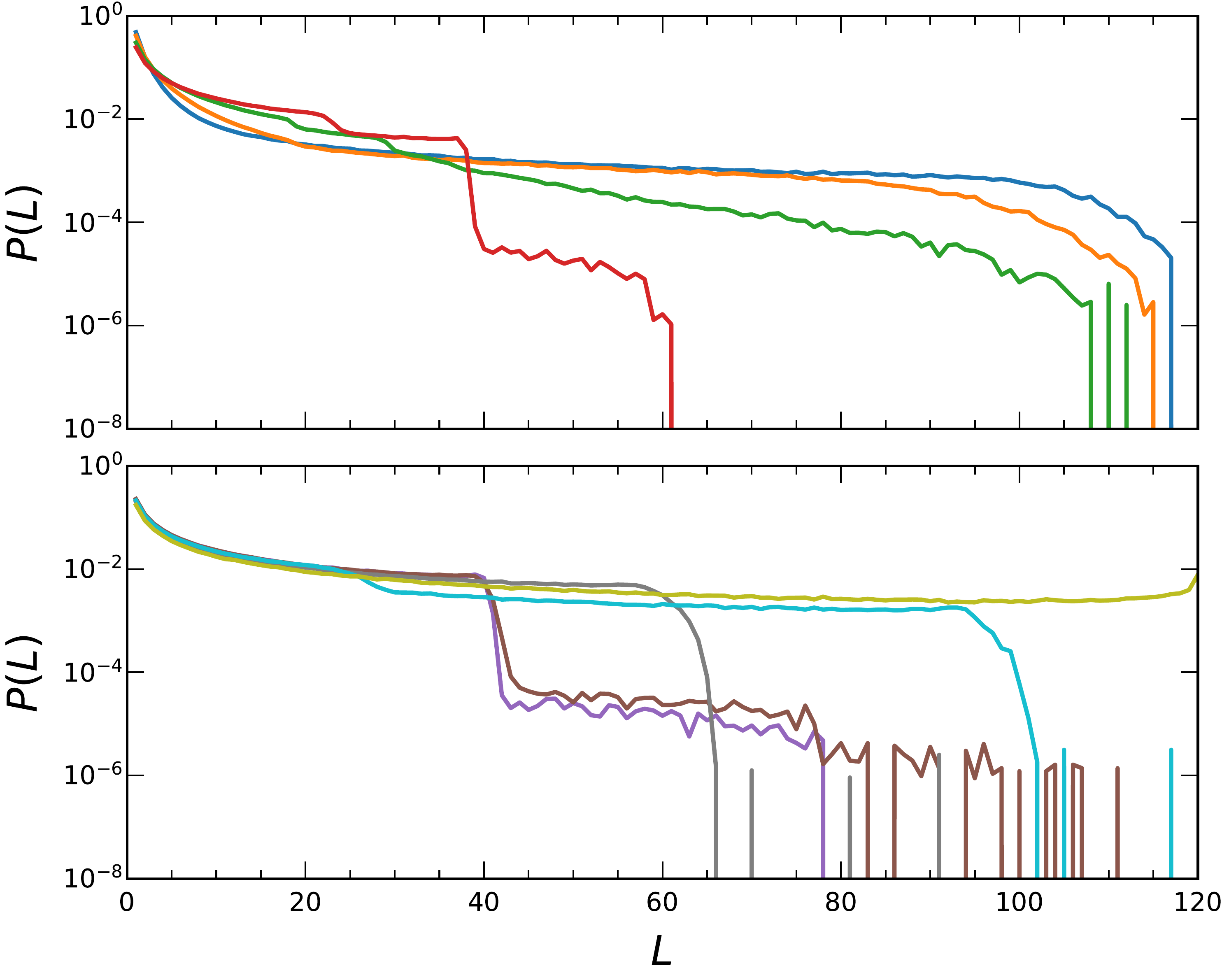}
\caption[]{Probability distribution $P(L)$ of the length of permutation cycles (semi-log scale) for increasing gravity.
Upper panel: $g=$ 0 (blue), 10 (orange), 20 (green), and 50 (red).
Lower panel: $g=$ 100 (purple), 150 (brown), 200 (grey), 210 (cyan), and 220 (olive green).}
\label{fig:perm_gu}
\end{figure}

In Fig.~\ref{fig:perm_gu} we report $P(L)$ for a number of $g$ values.
For $g=0$, a sizeable amount of polymers are formed by a number of world lines close to $N$, which explains the finite value of $f_s$.
In the absence of gravity there is a perfect, yet delicate equilibrium between the energetic advantage (lowering of potential energy) of having the polymers wound up inside the clusters and the ``entropic'' advantage (lowering of kinetic energy) determined by having particles delocalized --- i.e., polymers spread out --- over the entire surface.

However, already for $g=10$ we see a substantial depletion of large-$L$ cycles, which could be sufficient to suppress supersolidity altogether.
As $g$ grows further, permutation cycles get increasingly restricted to particles belonging to the same cluster, and this will likely cause the complete loss of coherence between the clusters.
Finally moving to $g=220$, the distribution of cycles develops a long tail and superfluidity would then be restored.

A more careful inspection of $P(L)$ in Fig.~\ref{fig:perm_gu} shows the presence of steps/jumps in most curves, which are initially tiny (e.g., for $g=10$), while becoming more spectacular --- at least, on a logarithmic scale --- in the interval from $g=100$ to 200, where they bring about the existence of a cutoff length in the distribution.
As already mentioned, the origin of these steps must be traced back to localization of the particles, implying that permutation cycles longer than the number of particles per cluster are statistically suppressed.

\begin{figure}[t]
\centering\includegraphics[width=\linewidth]{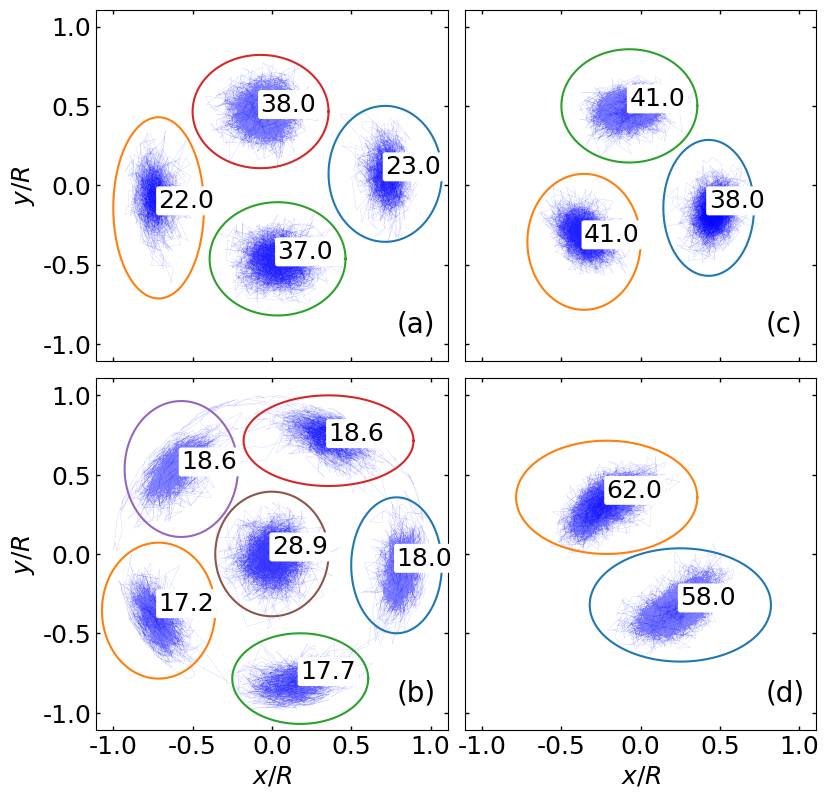}
\caption[]{Typical world line configurations. The number associated to each cluster is the mean number of particles contained in the ellipse (see text): (a) $g=20$, (b) 50, (c) 150, and (d) 200.}
\label{fig:ellissi}
\end{figure}

However, a more refined explanation is needed for $g=20$ and $g=50$, where two steps are actually seen in the profile of $P(L)$ (see upper panel of Fig.~\ref{fig:perm_gu}).
To clarify this point, in  Fig.~\ref{fig:ellissi} we have plotted the projections on the $x$-$y$ plane of the polymer beads in typical configurations for $g=20,50,150$, and 200.
The figure also shows colored ellipses which approximately delimit the clusters. We associate a number of bosons to each cluster by counting the number of world line beads inside the ellipse and dividing by $M$.

Except for unimportant fluctuations, clusters are nonequivalent for $g=20$ and 50, while being equally populated for $g=150$ and 200.
In the latter cases, the cluster size also represents the cutoff length in $P(L)$.
Instead, for $g=20$ each petal contains less particles than the cluster at the south pole.
The number of particles in a petal ($\approx 18$) compares well with the $L$ value at which the first step occurs in the histogram.
This corresponds to having a significant probability only for permutation cycles shorter than $L\approx 18$.
Likewise, the second step at $L\approx 30$ is related to the population of the central cluster:
cycles composed of more than $\approx 30$ world lines have a negligible probability.

For $g=50$ the interpretation is similar.
Now, the four clusters present are only equivalent in pairs (the more populated ones being those closer to the south pole).
The two sizes, $L\approx 22$ and $L\approx 38$, again correspond to the location of the steps in $P(L)$.

\begin{figure}[t]
\centering \includegraphics[width=\linewidth]{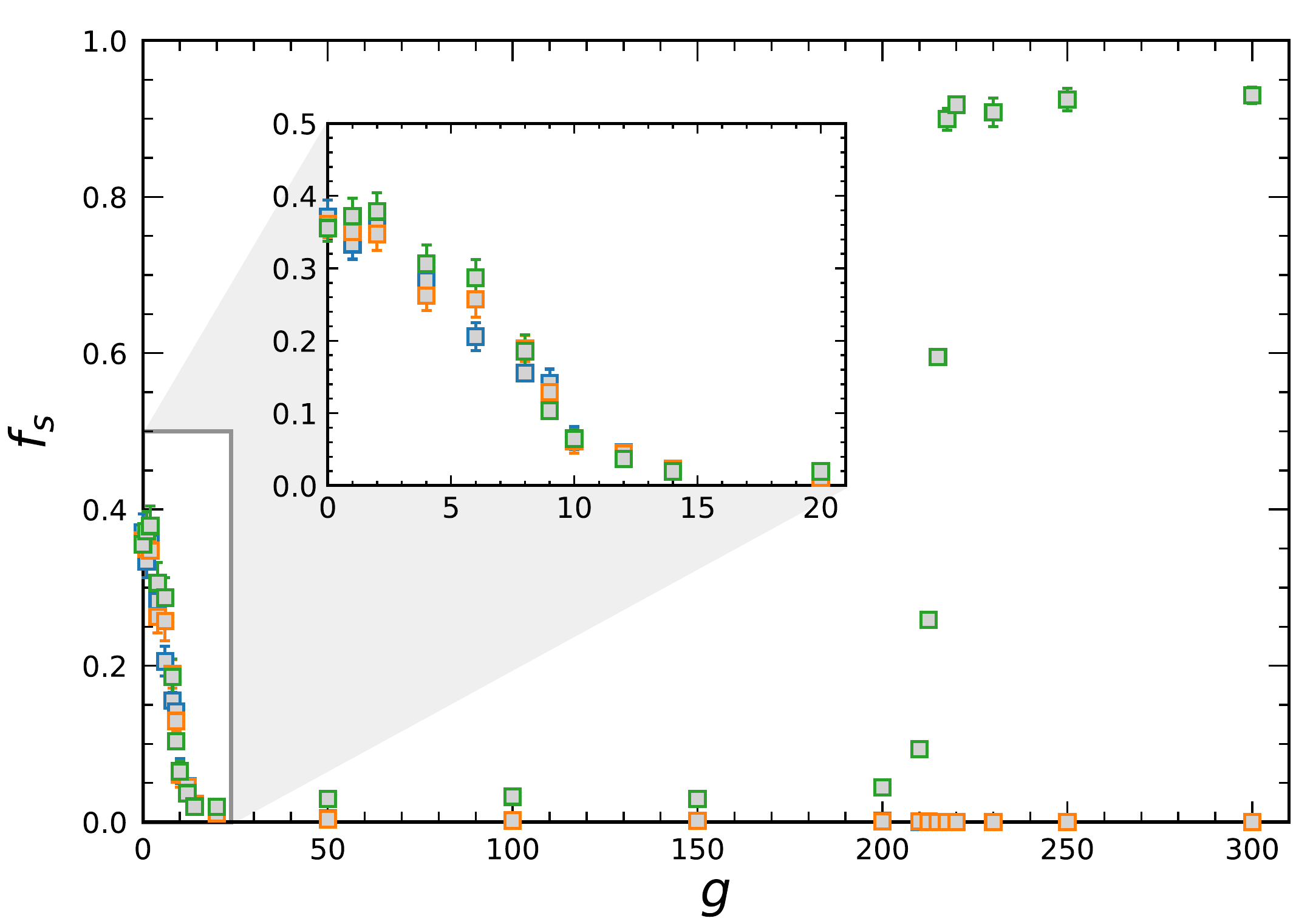}
\caption[]{Superfluid fraction of soft-core bosons on a sphere plotted as a function of gravity, along three spatial directions: $f_s^{(x)}$ (orange), $f_s^{(y)}$ (blue), $f_s^{(z)}$ (green).
The inset shows a magnification of the low-$g$ region.}
\label{fig:fs_vs_gu}
\end{figure}

\vspace{5mm}
As discussed in Section~\ref{method}, we can assess the quantum character of the system by computing the superfluid fraction $f_s$ from the simulation.
The values of $f_s$ along the $x,y$, and $z$ directions are plotted in Fig.~\ref{fig:fs_vs_gu} as a function of $g$.
Although the temperature was $T=0.2$, we have checked for a few $g$ values that $f_s^{(i)}$ remains practically constant at lower temperatures.
For this reason, Fig.~\ref{fig:fs_vs_gu} actually provides ground-state information.

Looking at the figure, we see three different regimes of gravity, i.e., $(a)$ $g\lesssim 10$, $(b)$ $10\lesssim g\lesssim 200$, and $(c)$ $g\gtrsim 210$.
For $g=0$, the superfluid fraction is non-zero thanks to the combination of a very small temperature (which keeps the world lines diffuse enough) and a well-tuned value of $\lambda$ (which gives quantum particles assembled in clusters the ability to contrast localization).
In this perfectly symmetric case, $f_s$ is the same along any direction.

In the whole $(a)$ region the quantities $f_s^{(i)}$ are the same and non-zero (to within statistical errors).
Thus, it seems reasonable to conclude that --- even though the original icosahedral structure is deformed by gravity --- the system is still supersolid.
However, as $g$ grows from 0 to 10 the superfluid fraction decreases, until it vanishes close to $g=10$.

\begin{figure}[t]
\centering \includegraphics[width=\linewidth]{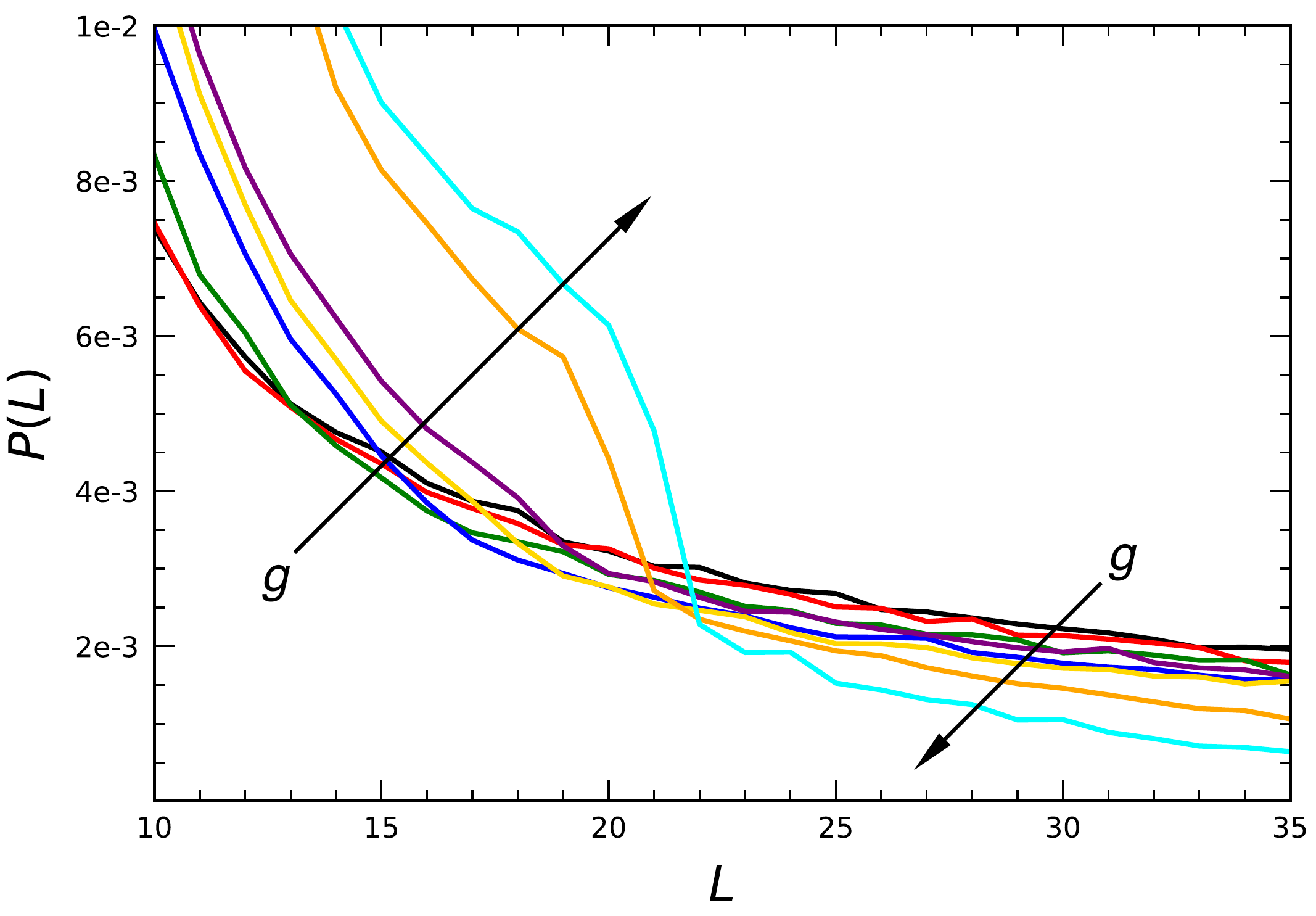}
\caption{Magnification of $P(L)$ in the low-$L$ region for a few values of $g$: 0 (black), 2 (red), 4 (green), 6 (blue), 8 (yellow), 10 (violet), 12 (orange), and 14 (cyan). The arrows indicate the direction of growing $g$.}
\label{fig:lowgperm}
\end{figure}

The behavior of $P(L)$ offers the key to read the $f_s$ data in the (a) region --- see Fig.~\ref{fig:lowgperm}, where the focus is on the statistics of permutation cycles in the small-$L$ range.
On increasing $g$, a step gradually arises in $P(L)$ near $L=N/6$, due to polymers wound up in a single cluster becoming more and more preferred to polymers distributed among various clusters.
When the occurrence of long cycles becomes too low, which happens for $g\approx 10$, $f_s$ falls down to zero.
The reason why long cycles are the more suppressed at stronger gravity calls upon the exclusion/interdiction of polymers from a surface region of increasingly larger area, making the benefits of delocalization progressively smaller.

\begin{figure}[b]
    \centering \includegraphics[width=\linewidth]{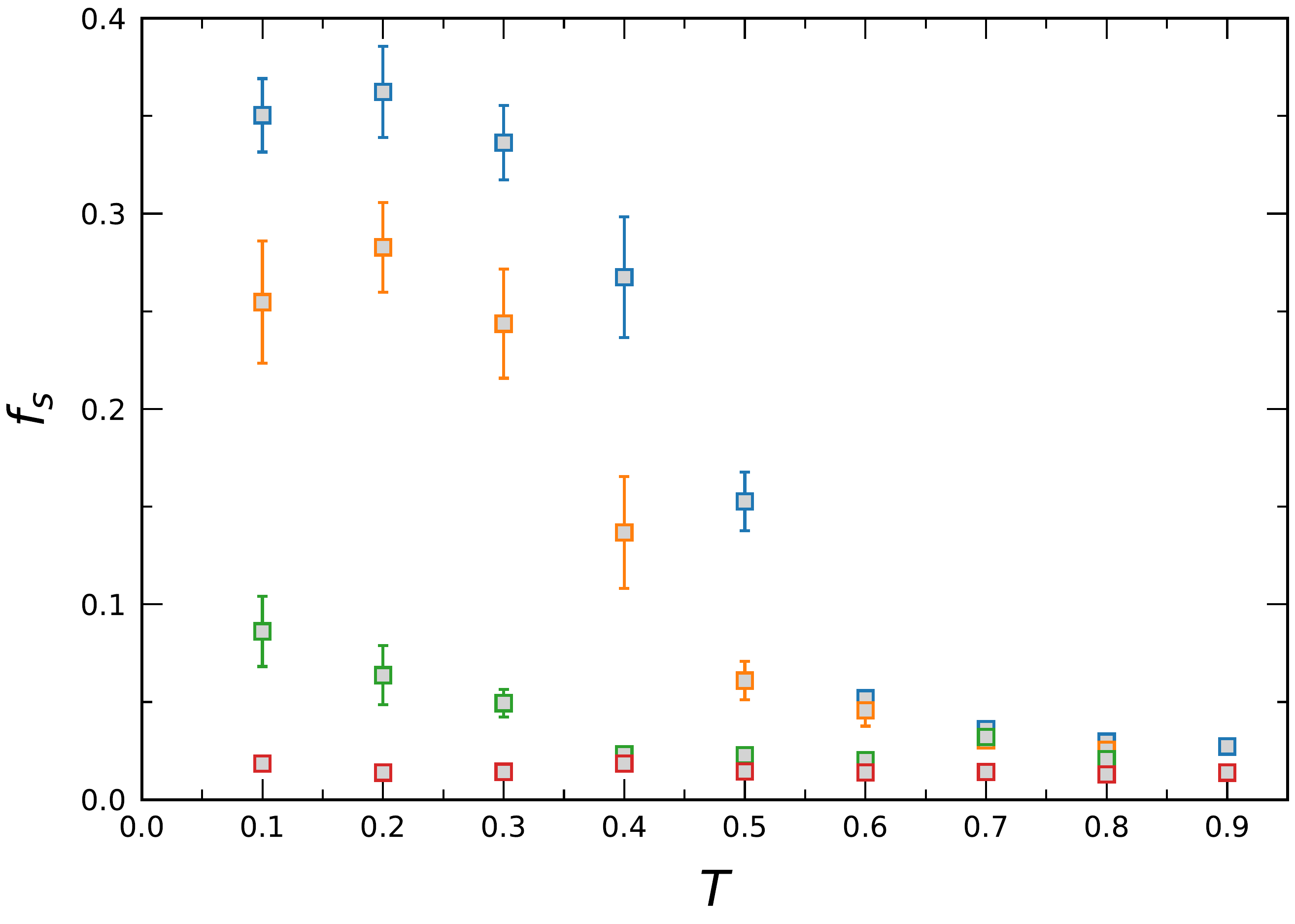}
    \caption[]{Superfluid fraction plotted as a function of temperature for a few values of $g$: $g=0$ (blue), 4 (orange), 10 (green), and 20 (red).}
    \label{fig:fs_vs_t}
\end{figure}

In the weak-gravity regime, we have also investigated the behavior of $f_s$ as a function of temperature, see Fig.~\ref{fig:fs_vs_t}.
We find that the temperature where $f_s$ vanishes is progressively reduced with increasing $g$.
In other words, the width of the (a) region shrinks when the temperature is increased.

In the (b) region (roughly, from $g=10$ to 200), gravity is sufficiently strong that the system does not support superfluidity any longer.
In particular, the quantities $f_s^{(x)}$ and $f_s^{(y)}$ are nearly zero.
In contrast, superfluidity along the field direction starts from zero at $g=10$ and then grows steadily with $g$, while remaining small anyway (probably, the non-zero value of $f_s^{(z)}$ is just a finite-size effect).

As a side comment, it is amazing that for a gravity as strong as $g=200$ the system still prefers to form two clusters rather than a single cluster at the pole.
This is due to the energetic advantage owned by a configuration with two clusters over one with a single cluster.

Beyond $g\approx 210$ (i.e., in the (c) region) gravity definitely overcomes the soft-core interaction and $f_s^{(z)}$ jumps to values close to 1 ($f_s^{(x)}$ and $f_s^{(y)}$ remain zero).
Anisotropy of the superfluid response is not new, since the same occurs for dipolar bosons on the sphere~\cite{ciardi2024}.
The resurgence of superfluidity in the (c) region happens in coincidence with the transition from two clusters to one.
When the clusters are two, their spatial separation is so high that permutation cycles longer than $N/2=60$ cannot occur, and $f_s^{(z)}$ vanishes.
If, instead, the system is all in one piece, then it can host permutation cycles of any length and becomes superfluid.

Finally, in order to ascertain the size dependence of our results, we have doubled the system size while keeping the density fixed.
By carrying out long PIMC simulations on a sphere with radius $\sqrt{2}$ times larger than before, we have verified that, in the low-$g$ regime, the superfluid fraction remains substantially unchanged relative to $N=120$.

\vspace{5mm}
Our analysis would not be complete if we did not make an attempt to estimate the effective value of $g$ for a system of Rydberg-dressed atoms subject to Earth gravity.

As mentioned above, we can take as an example a surface density of 4 $\mu$m$^{-2}$, and the mass of a Rubidium atom, $m \approx 87$ amu. In this case, vanishing superfluidity would occur at gravity acceleration around $a \approx$ 12 m/s$^2$, meaning that the atoms would still be supersolid in the presence of Earth gravity. For heavier bosonic atoms, superfluidity would vanish at smaller values of $a$. Details of the calculation are presented in Appendix~\ref{appc}.



\section{Conclusions} \label{conclusions}

In the study of the emergent properties of an ultracold gas of bubble-trapped atoms, the gravitational pull is a critical issue since it precludes a uniform distribution of atoms in the shell. We have explored whether this effect is milder in a strongly-interacting system  well beyond the dilute limit, taking the cluster supersolid of soft-core bosons on a sphere as our case study.

In the absence of gravity, the clusters are centered at the vertices of an icosahedron.
Using path integral Monte Carlo simulations, we track the evolution of the system structure as a function of the gravity strength $g$, assumed as a free parameter.
We thus find a sequence of transitions in the arrangement of clusters, which reduce in number as $g$ increases, until only one cluster is left at the bottom of the sphere.
In turn, also the superfluid response along the vertical axis is strongly affected by gravity, which makes it (nearly) vanishing already for $g\approx 10$ (in our units), until the final recovery when all particles condense at the bottom of the trap ($g>200$).
In this regard, the statistics of permutation cycles proves effective in unraveling the connection between cluster structure and superfluidity.
While the tail of the distribution informs on the extent of superfluidity in the system, its fine structure (jumps and cutoff length) carefully reflects the number and size dispersity of the clusters.

It turns out that quantum coherence is assured only when gravity is a weak perturbation (clusters are many and have fuzzy margins) or fully dominant (clusters ``merge'' together into one).
It appears that, depending on the atomic species involved, Earth gravity can be ``weak''; in this case, signatures of supersolidity of soft-core bosons in a spherical shell could be even seen in a laboratory on Earth.

We hope that the numerical evidence provided in this paper will stimulate experiments that explore strong-coupling regimes in bubble traps, in a similar way to what happened for supersolids in flat space, where theoretical and numerical results have long preceded implementations in the lab.\\

\begin{acknowledgments}

This work was supported by the European Union through the Next Generation EU funds through the Italian MUR National Recovery and Resilience Plan, Mission 4 Component 2 - Investment 1.4 - National Center for HPC, Big Data and Quantum Computing (CUP B83C22002830001). 
M. C. and F. C. acknowledge financial support from PNRR MUR Project No. PE0000023-NQSTI. 
FC and GP acknowledge the NICIS Centre for High Perfomance Computing, South Africa, for providing computational resources.

\end{acknowledgments}

\bibliography{mybose.bib}

%


\appendix
\section{Angular moves in PIMC}\label{appa}

As detailed in Refs.~\cite{ceperley1995} and \cite{boninsegni2006a}, the standard implementation of PIMC is based on the generation of world-line configurations drawn from a probability distribution corresponding to the free-particle density matrix. The proposed configurations are then accepted or rejected based on the strength of the potential terms.

In the standard PIMC formulation, the shape of a world line follows that of a free quantum particle at temperature $T$: a particle occupies a space proportional to the cube of the thermal wavelength $\Lambda = \sqrt{2 \pi \hbar^2 \beta / m}$. Clearly, in order to take the spherical constraint into account, the standard sampling procedure must be modified to ensure that the world lines entirely lie on the spherical surface.

In practice, segments of world lines with starting position 
$\textbf{r}_i$ and end position $\textbf{r}_f$ are constructed through a diffusion process which bears the name of Levy bridge. For each new bead $j$, with $j = i+1,\dots,f-1$, a temporary position $r_j^0$ is selected on a straight line from $\textbf{r}_{j-1}$ to $\textbf{r}_f$. Then, the temporary position is displaced by a vector, sampled from a Gaussian distribution derived from the free-particle density matrix $\rho_{free}$. Through these steps, the new position $\textbf{r}_j$ is generated and the process is repeated for the other beads. In the case of particles bound to a spherical surface, both steps need to be changed.

\begin{itemize}
\item Instead of centering the temporary position $r_j^0$ on a straight line from $\textbf{r}_{j-1}$ to $\textbf{r}_f$, we place it on the arc connecting $\textbf{r}_{j-1}$ to $\textbf{r}_f$ along the surface of the sphere.

\item Instead of drawing the displacement vector from a 3D Gaussian distribution of width $\Lambda$, we first draw a two-dimensional vector from a Gaussian distribution of width $\sigma_\theta$, which is usually close to $\Lambda$. The resulting vector $\textbf{v}$ must be converted to an angular displacement $\Delta \theta$ on the sphere.  
There are different ways to do this, so the specifics of the move can vary, but this is not important since detailed balance will take care that small differences in the generated paths are smoothed out. For example, we can generate the length of arcs along the sphere; in this case, we draw the vector $\textbf{l}_\theta$ from the two-dimensional Gaussian distribution. We use its modulus $l_{\theta} = |\textbf{l}_\theta|$ as the arc length, and set $\Delta \theta = l_{\theta} / R$. Another option is that $l_\theta$ is the length of a chord, in which case $\theta= 2 \arcsin\left(l_\theta/2R\right)$. We have tested both solutions and found no significant differences in the results. We use the first option in our algorithm. The probability to draw a specific choice of $\Delta \theta$ is then
\begin{equation} \label{eq:gaussian_sphere}
    P(\Delta r, \Delta \theta) = \frac{1}{2\pi\sigma_{\theta}^2} e^{-\frac{\Delta {\theta}^2 R^2}{2\sigma_{\theta}^2}}.
\end{equation}
\end{itemize}

Given the starting and final position, we find the normal unit vector $\hat{n} = \textbf{r}_{j-1} \times \textbf{r}_f / |\textbf{r}_{j-1} \times \textbf{r}_f| $ and the angle between them $\alpha = \textbf{r}_{j-1} \cdot \textbf{r}_f / |\textbf{r}_{j-1}| |\textbf{r}_f| $. Then, we make use of the definition of rotation matrix in $d=3$,
\begin{equation}
    R (\hat{n}, \alpha) = \begin{bmatrix}
    t n_x^2 + c & t n_x n_y - s n_z & t n_x n_zy + s n_y \\
    t n_x n_y + s n_z & t n_y^2 & t n_y n_z - s n_x \\
    t n_x n_zy - s n_y & t n_y n_z + s n_x & t n_z^2
    \end{bmatrix}
\end{equation}
where $c=\cos{\alpha}$, $s = \sin{\alpha}$, and $t = 1 - c$. Calling $m = f - (j-1)$, we then have that the temporary position $\textbf{r}_j^0$ is at
\begin{equation}
    \textbf{b} = R (\hat{n}, m \alpha) \, \textbf{r}_{j-1} + m \Delta r \frac{\textbf{r}_{j-1}}{|\textbf{r}_{j-1}|} \,.
\end{equation}

Once we have chosen the vector $\textbf{b}$, it is convenient to perform the calculation in a standard reference frame by bringing $\textbf{b}$ to coincide with the $z$ axis. We first compute the vector $\textbf{a} = \textbf{b} \times \hat{z} =  (b_y/b, -b_x/b, 0)$, then the rotation matrix $R (\textbf{a}/|\textbf{a}|, \arccos(b_z/b))$. This is the rotation matrix that brings $\textbf{b}$ to be aligned with $\hat{z}$, while $R^T$ is the matrix that restores the original vector.

Now, we generate a new position by drawing $\textbf{v}_\theta$ from \eqref{eq:gaussian_sphere}. The vector $\textbf{v}_\theta$ gives both a displacement and a direction in the $x$-$y$ plane, which we convert into coordinates $\theta, \phi$ on the sphere: $\theta = |\textbf{l}_\theta| / b$, $\phi = \arctan(l_\theta^y / l_\theta^x)$. The new vector is then simply
\begin{equation}
    \textbf{c} = b \left( \cos\phi \sin\theta \hat{x} + \sin\phi\sin\theta \hat{y} + \cos\theta \hat{z} \right) \,,
\end{equation}
and we go back to the original coordinate system by
\begin{equation}
    \textbf{r}_j = R (\textbf{a}/|\textbf{a}|, \arccos(b_z/b))^T \textbf{c}    ,.
\end{equation} 

The above procedure ensures that, when we build a Levy bridge as part of a PIMC move, the new beads always lie on the surface of the sphere. An analogous procedure can be applied to off-diagonal moves to construct open world lines.

Since we have changed the sampling distribution, the acceptance rates of the Monte Carlo moves must also change in order to satisfy the detailed balance condition. When going from configuration $s$ to $s'$, we must include a term of the form
\begin{equation}
    \frac{\prod_{j} P(\Delta \theta_{j,j+1})}{\prod_{j} P(\Delta \theta'_{j,j+1})} \frac{\prod_{j} \rho_{free}(\textbf{r}_j',\textbf{r}_{j+1}')}{\prod_{j} \rho_{free}(\textbf{r}_j,\textbf{r}_{j+1})}
\end{equation}
with $\Delta \theta_{j,j+1} = \textbf{r}_{j+1} \cdot \textbf{r}_j / |\textbf{r}_{j+1}| |\textbf{r}_j|$. \\

\section{Superfluidity and area estimator}\label{appb}

In bubble traps, $f_s$ is evaluated by sampling the so-called ``area estimator'' \cite{sindzingre1989,ciardi2022a}. This method draws a direct link between the area encompassed by connected polymers (describing the delocalization of bosonic particles) and the reduction in the moment of inertia compared to the classical case.
In this paper we examine the superfluid fraction along three orthogonal axes by computing, for $k=x,\,y$, and $z$,
\begin{equation}
\label{superf1}
f_{s}^{(k)} = \frac{4m^2}{\hbar^2\beta I_{\rm cl}^{(k)}}\left(\langle A_{k}^2 \rangle - \langle A_{k}\rangle^2\right)\,,
 \end{equation}
$A_{k}$ being the total area enclosed by particle paths projected onto the plane perpendicular to axis $k$.
In terms of particle positions $A_{k}$ reads: 
\begin{equation}
\label{superf2}
A_k =  \frac{1}{2} \sum_{i=1}^{N} \sum_{j=0}^{M-1} \left(\textbf{r}_{i,j} \times \textbf{r}_{i,j+1} \right)_k\,.
\end{equation}
Moreover, $I_{\rm cl}^{(k)}$ in Eq.~\eqref{superf1} is the classical moment of inertia relative to the $k$ axis. 
Finally, the observable in Fig.~\ref{fig:fs_vs_t} is defined as $f_s = (f_s^x + f_s^y +f_s^z)/3$.\\

\section{Estimate of $g$}\label{appc}

In our simulations, we use a unit of length $l$ and a unit of energy $E$ corresponding to physical quantities
$l_{\rm phys} $ and $E_{\rm phys} $. In the main text, these units are taken to be the soft-core radius $\sigma$ and the barrier height $\epsilon$.

To fix these values, we need to choose some physical parameters. For the length unit, we fix the density to about 4 $\mu$m$^{-2}$, as done in \cite{ciardi2024}. With $N=120$ and $R=1.4$, we then have
\begin{equation}
S_{\rm mks} = N / n_{\rm mks} = 30.0 \, \rm \mu m^2
\end{equation}
\begin{equation}
R_{\rm mks} = 1.545 \times 10^{-6} \, \rm m
\end{equation}
\begin{equation}
l_{\rm mks} = l R_{\rm mks} / R = 1.104 \times 10^{-6} \, \rm m
\end{equation}
Therefore $l_{\rm mks} \approx 1.1 \mu$m.

To define the unit of energy, we note that we can write for our value of $\lambda = \hbar^2 / 2m$ (set to $\lambda = 0.16$ in our simulation),

\begin{equation} \lambda / l^2 E = \lambda_{\rm mks} / l_{\rm mks}^2 E_{\rm mks} 
\end{equation}
Since $l$ and $E$ are 1 in the simulations, we write

\begin{equation} \label{eq:E_phys}
E_{\rm mks} = \lambda_{\rm mks} / \lambda l_{\rm mks}^2  
\end{equation}
We can calculate $\lambda_{\rm phys} = \hbar^2 / 2m$ using physical units. With $h = 6.63 \times 10^{-34}$ and choosing the mass of a Rubidium atom $m = 87 \times 1.67 \times 10^{-27}$, we find
$E_{\rm mks} \approx 1.96 \times 10^{-31}$ J.

In our simulations, we represent the gravitational potential as $V(z) = g z = m a z$. We need to establish the relationship between the value of $g$ in our simulations and that of $a_{\rm mks}$. We start from
\begin{equation} g l / E = a_{\rm mks} \, m_{\rm mks} \, l_{\rm mks}  / E_{\rm mks}\,. 
\end{equation}
We can replace $E_{\rm mks}$ with its expression in \eqref{eq:E_phys}, obtaining
\begin{equation} 
a_{\rm mks} = \frac{g \, \lambda_{\rm mks}}{m_{\rm mks} \, l_{\rm mks}^3 \, \lambda} = \frac{g \, \hbar^2}{2 \, m_{\rm mks}^2 \, l_{\rm mks}^3 \, \lambda} \,.
\end{equation}
Replacing $\lambda = 0.16$ and $l_{\rm mks} = 1.104 \times 10^{-6} \, \rm m$, we obtain:
\begin{equation} 
a_{\rm mks} = \frac{2.59 \times 10^{-50}}{m_{\rm mks}^2} g = \frac{2.59 \times 10^{-50}}{m_{\rm mks}^2} g \frac{\rm m \,kg^2}{\rm s^2}
\end{equation}
Introducing $m_{\rm amu} = m_{\rm mks} / 1.67 \times 10^{-27}\,\rm kg$, we can rewrite Eq.~(C8) as
\begin{equation} 
a_{\rm mks} = \frac{9270}{m_{\rm amu}^2} \, g \frac{\rm m}{\rm s^2} \,.
\end{equation}
For Rubidium, $m_{\rm amu} = 87$ so that
\begin{equation} 
a_{\rm mks} = 1.22 \, g \frac{\rm m}{\rm s^2} \,.
\end{equation}

Thus, we see that a physical value of $a_{\rm mks} = 9.8 \, \rm m / s^2$ corresponds to a value of $g \approx 8$ in our simulations.

\end{document}